\documentclass[aps, prd, twocolumn, lengthcheck, superscriptaddress, 
nofootinbib]{revtex4-1}

\usepackage{amsmath,amssymb,amsfonts}
\usepackage{slashed}
\usepackage{graphicx}
\usepackage{dcolumn}
\usepackage{bm}
\usepackage{tikz}

\def\bi{\bibitem}

\def\la{\langle}\def\ra{\rangle}
\def\be{\begin{eqnarray}}\def\ee{\end{eqnarray}}
\def\lsim{\mathrel{\rlap{\lower3pt\hbox{\hskip1pt$\sim$}}
		\raise1pt\hbox{$<$}}} 
\def\gsim{\mathrel{\rlap{\lower3pt\hbox{\hskip1pt$\sim$}}
		\raise1pt\hbox{$>$}}} 

\def\gsi{\rm GSI}
\def\riken{\rm RIKEN}
\def\FR{\it FR}

\begin{document}

\preprint{APS/123-QED}

\title{Probing for an IR-fixed Point in QCD \\ 
by Superallowed Gamow-Teller Transitions in Doubly Magic Nuclei
}


\author{Mannque Rho}
\email{mannque.rho@ipht.fr}
\affiliation{Universit\'e Paris-Saclay, CNRS, CEA, Institut de Physique Th\'eorique, 91191, Gif-sur-Yvette, France }

\author{Long-Qi Shao}
\email{shaolongqi22@mails.ucas.ac.cn}
\affiliation{School of Fundamental Physics and Mathematical Sciences, Hangzhou Institute for Advanced Study, UCAS, Hangzhou, 310024, China}
\affiliation{Institute of Theoretical Physics, Chinese Academy of Sciences, Beijing 100190, China}
\affiliation{University of Chinese Academy of Sciences, Beijing 100049, China}

\vskip 0.5cm

\date{\today}

\begin{abstract}
This brief note is to point out that the recent measurements at GSI and RIKEN of the superallowed Gamow-Teller (GT) transition in the doubly magic closed-shell nucleus $^{100}$Sn could give an  indication for a possibly important {\it fundamental} quenching, thus-far unrecognized, of $g_A$, unambiguously distinct from nuclear correlation effects in the framework of nuclear effective field theory. The result, either confirmed or ruled out both by experiment and by theory,  can have strong impacts on nuclear physics vis-\`a-vis with nuclear effective field theory as well as on particle physics  relevant for going beyond the Standard Model.
\end{abstract}

\maketitle


\subsection*{ GSI Data vs. RIKEN Data:\\ A Fundamental Issue?}
Two recent measurements of the superallowed Gamow-Teller transition in the doubly magic closed-shell (DMCS) nucleus $^{100}$Sn, one at GSI~\cite{GSI} and the other at RIKEN~\cite{RIKEN}, at first sight, give significantly different GT strengths. Put in the notation used in the two papers, the measured results are
 $ {\cal B}_{\rm GT}^{\gsi}   =9.1^{+3.0}_{-2.6}$ and ${\cal B}_{\rm GT}^{\riken}=4.4^{+0.9}_{-0.7}$. In terms of the ``extreme single-particle shell model (ESPM)" considered applicable for the DMCS nucleus $^{100}$Sn,  the quenching factor denoted in the literature as $q$  
\be
g^\ast_A=q g_A 
\ee
where $g_A=1.276$ is the value for the free-space neutron beta decay and $g_A^\ast$ is the ``experimentally determined" (central) value  in the ESPM\footnote{We avoid quoting the error bars which are difficult to interpret given the theoretical arguments injected in the quoted results, the reliability of which is hard to quantify in our approach. For the reason explained below, we will take the GSI value to be very close to $q^{\gsi}\approx  0.79$ giving $g_A^\ast\approx 1$ for GSI.} is found to be 
\be
q^{\gsi}\approx 0.73, \ q^{\riken}\approx 0.50.\label{issue}
\ee

The objective of this brief note is to point out, reiterating what has already been reported in the literature,  that this $\sim$ (30 - 40)\%, if the difference is not swamped in the error bars,  could be extremely important for both nuclear and particle physics.  Such a quenching could  involve a {\it fundamental renormalization (FR)}  -- as opposed to the usual (mundane) nuclear correlation effects including exchange currents --  of the axial coupling constant in the context of nuclear effective field theory ($n$EFT) for capturing QCD in the framework of Weinberg's ``Folk Theorem" on quantum effective field theory~\cite{FT,bira}. It could lead  via \cite{mr91}  to  a novel development discussed in \cite{MRgA,MRanomaly} by affecting nuclear processes where the effective $g^\ast_A$ figures in a ``fundamental" way.  Up to date, no nuclear many-body theory, ab initio or otherwise  heralded as ``first-principles," approaches have addressed this and related issues in nuclear physics. Perhaps most significant of all, it could impact on neutrinoless double beta decay in heavy nuclei being extensively studied experimentally for going beyond the Standard Model (BSM). There, with $g^\ast_A$ figuring in the decay rate $\propto (g^\ast_A)^4$,  the process would be  suppressed by as much as or even more than an order of magnitude.  It would also ruin what could be classed as the precision calculations of certain nuclear axial responses such as the axial-charge transitions in nuclei~\cite{KR,Warburton,minamisono}. More notable for QCD as a gauge theory is the possible presence of an IR fixed point for hidden scale (or conformal) symmetry which has been argued to be absent for the flavor numbers $N_f =2, 3$ -- a long-standing controversy --  specially relevant for nuclear physics. In fact there are, in the literature,  certain theoretical arguments that go even to the extent that such IR fixed points are to be  ruled out for small number of flavors.  

It should be fair to state that up to date, apart from the RIKEN data, there is actually no clear indication for the presence of or hint for an {\FR}  in the GT processes, particularly in light nuclei, say for $A <10$, where the chiral-symmetry-implemented nuclear field theory referred to as  $\chi$EFT (\`a la Weinberg's FT)~\cite{bira} has been treated in powerful Monte Carlo approaches to verify $q\approx 1$, i.e., no FR~\cite{wiringa}.  Meson-exchange currents enter in $\chi$EFT only at high chiral order, N$^m$LO, starting at $m= 3$, in nuclear processes involving the space component of the axial current as first pointed out in \cite{mr91,tsp}, so unless the leading-order (LO) single-particle GT operator is accidentally suppressed -- i.e., such as by symmetry -- many-body currents {\it should not enter} at greater than a few \%  in the GT amplitude. This is indeed what is found in \cite{wiringa}.\footnote{If for some particular kinematic or symmetry  reasons the N$^3$LO turnes out to be non-negligible, then there are no reasons (in the sense of the FT) why the next-order,  that is, N$^4$LO, terms can be ignored as was done in \cite{firstprinciples}.  The difficulty here is the practical -- even if not in-principle -- difficulty that  there are too many unknown parameters to fix at that order, so the given $\chi$EFT, as formulated,  begins to make no good sense~\cite{MRgA}.} Thus $g_A$ remains unaffected by {\it fundamental} renormalization in light nuclei. 

In heavier nuclei, however, there are various indications/hints that the quenching factor could be $q < 1$~\cite{Suhonen}.  But one cannot reliably address what the cause of the quenching could be. This is because it is not feasible to {\it separate  what's fundamental from what's not} due to the closeness of the relevant scales. Up to date,  there is no reliably controlled many-body technique in $\chi$EFT (anchored on only the nucleon and pion degrees of freedom) to capture the {\it full} nuclear correlations and compute what might  be categorized as the wave-function effect~\cite{wilkinson}, that is mundanely nuclear.    In particular, the GT transitions, particularly the superallowed ones, are extremely sensitive to the nuclear tensor force, including $\Delta$-hole components.  In heavier nuclei, with increasing density, the interplay between the pion and hidden local symmetry vector bosons ($\rho$ and $\omega$) and hidden scale symmetry boson (dilaton to be denoted as $D$ in this paper) integrated out into higher derivative terms in $\chi$EFT becomes extremely intricate at increasing density, bringing in certain important effects missed in $\chi$EFT such as in the tensor force. For instance put in what's called ``double-decimation" approach in Fermi-liquid theory of nuclear matter~\cite{DD-BR}, the tensor force turns out to be RG-invariant in nuclear medium~\cite{Pristine-Signal}. These features have been missing in current many-body calculations.
\subsection*{Resolving the Issue}
It has been suggested that this conundrum can be bypassed for the doubly-magic shell nucleus $^{100}$Sn. Here one can do a mapping between the Landau(-Migdal)\footnote{From here on we will think in terms of Landau-Migdal given that the pion field enters in the theory.}  Fermi-liquid model based on renormalization-group approximation on nucleons on the Fermi sphere at the Fermi liquid  fixed-point (FLFP)~\cite{shankar} and the extreme single-particle shell model (ESPM) in doubly magic closed shell (DMCS)  nucleus here,  $^{100}$Sn~\cite{GSI}. This is to combine $\chi$EFT (with BR scaling heavier degrees of freedom implemented)  and renormalization-group for the strongly correlated fermions on Fermi surface~\cite{shankar}. The details are given in \cite{MRgA,MRanomaly}. Very briefly restated,  the idea is that the superallowed GT transition in the DMCS nucleus can be looked at as a quasiparticle on the Fermi surface making the superallowed $q/\omega\to 0$ transition on the Fermi surface. The FLFP corresponds to the limit that $1/\bar{N}\to 0$ where $\bar{N}=k_F/(\Lambda_{FL}-k_F)$ where $\Lambda_{FL}$ is the cutoff above the Fermi sea. This limit corresponds to the suppression of the quasi-particle-quasi-hole loops. In terms of the hidden symmetry -- both hidden local and hidden scale -- implemented chiral Lagrangian ${\cal L}_{{\rm HS}\chi}$,  this limit can be simulated in the DMCS system by making the shell-model transition from the ground state of $^{100}$Sn to the {\it lowest single} neutron-particle-proton-hole state in the daughter nucleus  $^{100}$In with the neighboring  p-h states ignored.  Now measuring the transition to the lowest state requires an experimental finesse and how well this feat can be achieved is a question to be further scrutinized. Ref. \cite{GSI} claims that this can be done at 95\%,  making $q^{\gsi}$ closer to 1. We will take this estimate in our numerical consideration. 

One important question that should be answered  in making the mapping of the two approaches mentioned above remains more or less unanswered.  One attempt to address this question was to apply  the method of coadjoint orbits developed recently  in condensed matter theory~\cite{coadjoint}.  It is to render a systematic power expansion of  the Fermi-surface fluctuations going beyond the large $\bar{N}$ approximation replacing the chiral power counting in $\chi$EFT.  Assuming the chiral symmetry properties of pionic mode are of the same character as those found in  \cite{FR}, we treated only the modes corresponding to  heavier degrees of freedom brought in by hidden symmetries.  We found what we consider as  $1/\bar{N}$ corrections to the FLFP approximation in the problem to be surprisingly small~\cite{shao-rho}. This is consistent with what one expects at the dilaton-limit fixed point, $g_A\to 1$ and $f_\pi\to f_\chi$, at densities relevant to the core of massive stars.   
\subsection*{Where does $q$ come from?}
The question as to where the quenching factor $q$ comes from has been a long-standing unanswered problem in nuclear physics~\cite{wilkinson,Suhonen}.  From the point of view of QCD, it could also implicate a long-standing issue as to whether QCD at the flavor number $N_f << 18$ has an IR fixed point~\cite{Crewther,crewther-tunstal,pion-zwicky,Zwicky}.\footnote{Crewther's ``genuine dilaton (GD)"~\cite{Crewther} scenario and Zwicky's ``QCD-conformal dilaton (QCD-CD)" scenario~\cite{Zwicky}, although most  likely different in detail, do however share the common structure of the IR fixed point populated by massless Nambu-Goldstone bosons and massive matter fields. For our discussion, we think their difference to be unimportant.}  We are considering the discrepancy between the GSI and RIKEN data, if real, as  zero-ing on this issue.

In our scheme as formulated in \cite{MRgA,MRanomaly}, the quenching factor $q$, in the leading chiral order (LO) and the scale dimension $d_s > 4$, is given by a product of two density-dependent factors 
\be
q(n) =q_{ssb} (n) q_{snc}(n)
\ee
where  $q_{ssb}$ encodes the quenching due to the scale anomaly associated with the IR fixed point and $q_{snc}$ (with ``snc" standing for strong nuclear correlations) is to capture  the (in-principle) complete nuclear correlation effects coming from nuclear effective field theory.  At higher orders in the combined chiral-scale  expansion, $q$ may have corrections to the simple product form. In the past development~\cite{Suhonen}, $q_{ssb}$ has been taken to be 1, ignoring the  possible effect of scale anomaly. Our claim is that it is in the RIKEN result (\ref{issue}) that the possibility $q_{ssb} < 1$ is first brought up.

Calculating $q(n)$ in $^{100}$Sn with appropriate caveats was already reported in \cite{MRgA}.  There $q_{snc}$ was computed in nuclear effective field theory formulated  in the renormalization-group approach to strongly correlated fermions on the Fermi surface~\cite{shankar} starting from  a scale-invariant hidden local symmetric chiral Lagrangian~\cite{FR}.  (This approach has been referred to as G$n$EFT to be distinguished from the standard $\chi$EFT.)  Briefly restated, the idea~\cite{FR} was to to map the  hidden local symmetric  chiral Lagrangian made scale-symmetric by the ``conformal compensator field" $\chi=f_\chi e^{D}/f_\chi$ denoted as ${\cal L}_{\chi {\rm HLS}}$ to Fermi-liquid theory taking into account the BR scaling $\Phi(n)=f^\ast_\chi/f_\chi\simeq f^\ast_\pi/f_\pi$~\cite{br91}. One then takes the mean fields of ${\cal L}_{\chi{\rm HLS}} $ and identify them with the Fermi-liquid fixed-point quantities.  The formulation gives an excellent description of the ground-state quasiparticle properties and nuclear EW response functions in a simplified form~\cite{FR} when  the quasiparticle interactions are expressed in terms of the Fermi-liquid fixed-point Landau interaction parameters {\it that are BR-scaling}.  {\it That the BR scaling figures in the formulation renders the Landau Fermi-liquid fixed-point approximation more effective for the problem.} It has been further improved in G$n$EFT to handle even higher densities capturing the quark-hadron continuity in the density regime applicable to compact stars~\cite{GnEFT}.  This procedure gives, in the large $N_c$ limit and the FLFP approximation, the deceptively simple formula for the (``snc") quenching factor\footnote{This can be interpreted as an in-medium Goldberger-Treiman relation with a validity roughly of  the vacuum relation.} 
\be
q^L_{snc}=(1-\frac 13\Phi \tilde{F_1^\pi})^{-2}\label{qsnc}
\ee 
 where $\tilde{F_1^\pi}=F_1^\pi (m_N/m^L_N)$ with $F_1^\pi$  the Landau parameter for the pion exchange -- absent in condensed matter systems -- and $m^L$ the Landau effective mass of the nucleon.  One can understand Equation (\ref{qsnc}) as a big improvement over Walecka's  linear mean-field theory brought in non-perturbatively from BR scaling. How this can be derived in terms of higher-order terms in scale-chiral power counting has not been yet fully worked out. But the presence of the BR scaling $\Phi$ and the connection $f_\pi\approx f_\chi$ -- which seems to be associated with the presence of an IR fixed point in QCD of Ref.~\cite{Crewther,Zwicky} which differs drastically from the case of large $N_f$~\cite{appelquist}-- reflects that the hidden local and scale symmetry that captures  the {\it complete} nuclear correlations within the given EFT describing the quasiparticle making the superallpwed GT transition on the Fermi surface. $\Phi$ is known from deeply bound pionic atoms up to $n\simeq n_0\approx 0.16\ {\rm fm}^{-3}$ and $F^\pi_1$ is entirely given by chiral symmetry with BR scaling implemented, so one gets immediately from (\ref{qsnc})
 \be
q^L_{snc}= 0.79 \ {\rm\  for}\  n=n_0 \label{qL}
\ee
giving 
\be 
g_A^L=1.276 \times 0.79\simeq 1.0.\label{gAL}
\ee  
This is essentially what the GSI measurement gave.  It can be interpreted with  $q_{ssb}\approx 1$, i.e., no fundamental quenching.
Since as found in \cite{MRgA}, $q^L_{snc} (n)$ as computed in the Fermi-liquid fixed theory is very little dependent on density, one can say this is also what's found in light nuclei with no fundamental quenching.  Recent shell-model calculations~\cite{sdshell} of the $s$-$d$ shell nuclei in the mass range of $A=18-39$ obtain $q = 0.794 \pm 0.05$ and  $0.815\pm 0.04 $ agreeing precisely with (\ref{qL}) which was first predicted in 1996~\cite{FR}. This again suggests that $q_{ssb}\approx 1$.
\subsection*{IR fixed point and fundamental quenching of $g_A$}
Now applying the same reasoning to the RIKEN data, one gets 
\be
g_A^{\ast{\rm riken}} \approx 1.276\times 0.79\times 0.6\approx 0.6.
\ee
This $\sim (30 - 40)\%$ reduction in $g_A$ due to fundamental renormalization of the axial coupling constant found in the ``improved" measurement, if real, could be the {\it serious problem}. It is noteworthy that this reduction, hidden in the matter-free vacuum, gets revealed in nuclear medium and should apply to $g_A$ in nuclear medium independently of where that constant appears, whether in GT or forbidden transitions or $\mu$ capture etc. and specially in fourth power in neutrinoless double beta decays in heavy nuclei, affecting the decay rate by an order of magnitude. 

Up to here, there is  nothing new from the results already given in  \cite{MRgA}. 

What's new  in this paper is that $q_{ssb}$ now could be estimated theoretically from the anomalous dimension $\beta^\prime$ of the gluon energy-momentum tensor if  the IR fixed-point argument of \cite{Zwicky} were  adopted.

The basic idea is that applying the Callan-Symanzik's renormalization-group invariance to the nucleon-weak field amplitude,  gives, to the {\it leading chiral-scale order},  the coefficient of the  nucleon axial-current coupling~\cite{Crewther}.  In medium it is given by
\be
J^a_{5,\mu}=g_A \kappa {\bar{\psi} (\frac 12\tau^a \gamma_\mu\gamma_5)\psi }  
\ee
where 
\be
\kappa=  c_A + (1- c_A)({\chi}/{f_\chi})^{\beta_{IR}^\prime}\label{kappa}
\ee
where $c_A$ is a scale-dimension zero constant and $\beta_{IR}^\prime$ is the anomalous dimension of the gluon energy-momentum tensor
\be
-\frac{d}{d\ln \mu}  \ln G^2\equiv\gamma_{G^2}=\beta^\prime - \frac{\beta}{g}.
\ee
In medium $\chi\to f_\chi^\ast + \chi^\prime$ where $\ast$ stands for the density dependence, so that to the order considered $\kappa$ contributes a density-dependent renormalization of $g_A$
\be
q_{ssb}=c_A + (1- c_A)({f_\chi^\ast}/{f_\chi})^{\beta_{IR}^\prime}.\label{ssb}
\ee
If $\beta_{IR}^\prime=0$ (in or outside of nuclear medium) or in the matter-free space with $f_\chi^\ast=f_\chi$,  $q_{ssb}\to 1$, so the anomaly effect, if non-zero, will be  hidden in loops coming at higher scale order.  But in nuclear medium, if $\beta_{IR}^\prime\ne 0$, it gives a tree-level contribution to $q_{ssb}$.   For the IR fixed-point concerned~\cite{Crewther,Zwicky}, $f^\ast_\chi \approx f^\ast_\pi$ \footnote{The equality  is shown to hold at the Fermi-liquid fixed point. It also holds at the dilaton limit fixed point~\cite{vankolck}  in G$n$EFT~\cite{GnEFT}. Therefore it is fair to assume that it holds in nuclear matter.} and $f_\pi^\ast$ is known experimentally up to $n =n_0$. At present, there is no theoretical information on the constant $c_A$ and no lattice data for $\beta^\prime$.  Therefore $q_{ssb}$ cannot be determined theoretically. If one were to obtain data on the superallowed GT transition of doubly magic closed shell nuclei at two different densities, both $c_A$ and $\beta^\prime$ could perhaps be fixed given that they will be insensitive to density dependence.  Note that most intriguingly,  {\it it is the density of the matter that would expose and make visible $q_{ssb}$ hidden in the dilaton loops or the vacuum modified by nuclear density.}

Now here is the key point  of this paper that we believe deserves to be stressed: A possible structure of  $\beta_{IR}^\prime$ is given by  Zwicky's theory~\cite{Zwicky} and it is that the non-vanishing quark condensate breaks {\it both} the chiral and the scale (or conformal) symmetry spontaneously.  The role of the strange-quark mass that figures in the IR fixed-point  in the schemes of Refs.\cite{Crewther,Zwicky} may therefore have to be much better understood for more accurate prediction. 

As it stands, however, the RG and EFT methods in the IR fixed-point scheme of \cite{Zwicky} make a simple prediction\footnote{This prediction was inspired by its analogy to ${\cal N}=1$ supersymmetric gauge theories~\cite{susy} and considered to be applicable to the nonsupersymmetric case considered here.}   
\be
\beta_{IR}^\prime=0.
\ee  
This implies that $\kappa\to 1$ and hence
\be
q_{ssb}\to 1.
\ee

\subsection*{Concluding Remarks}
The recent measurement at RIKEN of the superallowed Gamow-Teller transition in the doubly magic closed-shell nucleus $^{100}$Sn is found to require $\sim (30-40) \%$  {\it fundamental} quenching, i.e., $q_{ssb} \sim (0.6 - 0.7)$, associated with the conformal anomaly distinct from strong nuclear correlation effects, $q_{snc}^L=0.79$,  captured  by the Landau Fermi-liquid fixed-point approximation.   {\it This {fundamental} effect is missing in \cite{firstprinciples}}. It  appears however to be in disagreement  with the current scenario~\cite{Crewther,Zwicky} of the scale (or conformal) symmetry breaking in QCD of  the flavor number $N_f\leq 3$ with the presence of an IR fixed point. The Crewther-Zwicky scenario, interpreted in the G$n$EFT mapped to the ESPM and applied to $^{100}$Sn in the leading order scale-chiral symmetry expansion, predicts  $q_{ssb} \sim 1$, a negligible fundamental quenching. This apparently is  in support of the older GSI result which is  also seemingly consistent with the presently available precision nuclear axial processes such as the axial-charge transitions and GT transitions in light nuclei treated in simple shell-model.  Could this be because the RIKEN result is in error or  the $\beta_{IR}^\prime$ in Eq.~(\ref{ssb}) should receive large corrections? The log dependence of the $\beta$ function near the IR fixed point argued in \cite{Zwicky}, associated with the role of the strange-quark mass for inducing spontaneous scale-symmetry breaking, may need to be worked out to answer this question~\cite{shaowork}.

There are several important issues raised in the conundrum involved here.   Given that the RIKEN data is, as argued by the authors of \cite{RIKEN}, likely a big improvement over the GSI data and since the issue raises highly crucial point -- which seems to remain unrecognized in the field,  revisiting this process,  both experimentally and theoretically, is in high priority order. The ab initio  ``first-principles" approach~\cite{firstprinciples} must have clearly failed to capture the $q_{ssb}$ of the RIKEN data if it were proven to be correct.   That raises the question as to  how the anomalous dimension $\beta^\prime_{IR}$ can be implemented in the ab initio approach heralded as the ``first-principles solution" to the problem. 
\subsection*{Acknowlegments}
We are grateful to Roman Zwicky for very useful comments and suggestions on the approach to the IR fixed point in hidden conformal symmetry in QCD.

\end{document}